# Explorability and the origin of Network Sparsity in Living Systems

Daniel M. Busiello[1], Samir Suweis[1], Jorge Hidalgo[1] and Amos Maritan[1]

[1] Department of Physics and Astronomy, University of Padova, CNISM and INFN, 35131, Padova Italy

**The increasing volume of ecologically and biologically relevant data has revealed a wide collection of emergent patterns in living systems. Analyzing different datasets, ranging from metabolic gene-regulatory to species interaction networks, we find that these networks are sparse, i.e. the percentage of the active interactions scales inversely proportional to the system size. This puzzling characteristic has been neither yet considered nor explained. Herein, we introduce the new concept of explorability, a measure of the ability of the system to adapt to newly intervening changes. We show that sparsity is an emergent property resulting from a variational principle aiming at the optimization of both explorability and dynamical robustness, the capacity of the system to remain stable after perturbations of the underlying dynamics. Networks with higher connectivities lead to an incremental difficulty to find better values for both the explorability and dynamical robustness, associated with the fine-tuning of the newly added interactions. A relevant characteristic of our solution is its scale invariance, that is, it remains optimal when several communities are assembled toghether. Connectivity is also a key ingredient determining ecosystem stability and our proposed solution contributes to solving May's celebrated complexity-stability paradox.**

In inanimate matter, elementary units, such as spins or particles, always have their mutual interactions turned on (with an intensity decaying with their relative distance), and thus the interaction network is dense, with all connections present, i.e. particles do not have the freedom to adjust or change their interactions, unless they change their relative distances. In contrast, living systems are composed of interacting entities, such as gene [1, 2, 3, 4], metabolites [1, 5, 6], individuals [7, 8, 9] and species [4, 10, 11, 12, 13, 14], with the ability to rearrange and tune their own interactions in order to achieve a desired output [1]. Indeed, thanks to advances in experimental techniques, which are generating an increasing volume of publicly available ecologically and biologically relevant data, several studies indicate that interaction networks in living systems possess a non-random architecture characterized by the emergence of recurrent patterns and regularities [10, 11, 15, 16].

Analyzing different dataset of ecological, gene-regulatory, metabolic and other biological interaction networks (see Supplementary Information and references therein), we find that one ubiquitous pattern is sparsity [1, 2, 11, 12, 17, 18], i.e. the percentage of the active interactions (connectivity) scales inversely proportional to the system size (illustrated in Fig. 1). For example, in the case of ecological systems, species interact selectively even when they coexist at short distances and most of the interactions are turned off. A generic system formed by N interacting units may have a maximum number of interaction equal to $N(N-1)/2 \sim N^2$, i.e a connectivity, $C$, equal to 1. On the other hand the minimum number of interactions that guarantees that the interaction network is connected is of order $N$, that is $C \sim 1/N$, corresponding to the percolation threshold of random networks [19]. Thus, in this range of possible connectivities, it is quite surprising that the observed ones in the interaction networks of many living systems correspond to the lowest possible values. However, it is not known if this recurrent property gives any advantage or reward to the system, and a theoretical framework to understand the origin of sparsity is still lacking.

In this work we propose a variational approach [6] to describe the role of active interactions in living systems, and we show that sparse networks offer, at the same time, a maximum capability of the system to visit as many stable attractors as possible by simply tuning the interaction strengths (explorability) as well as the largest robustness of the underlying dynamics, guaranteeing that such attractors remain stable (dynamical robustness).

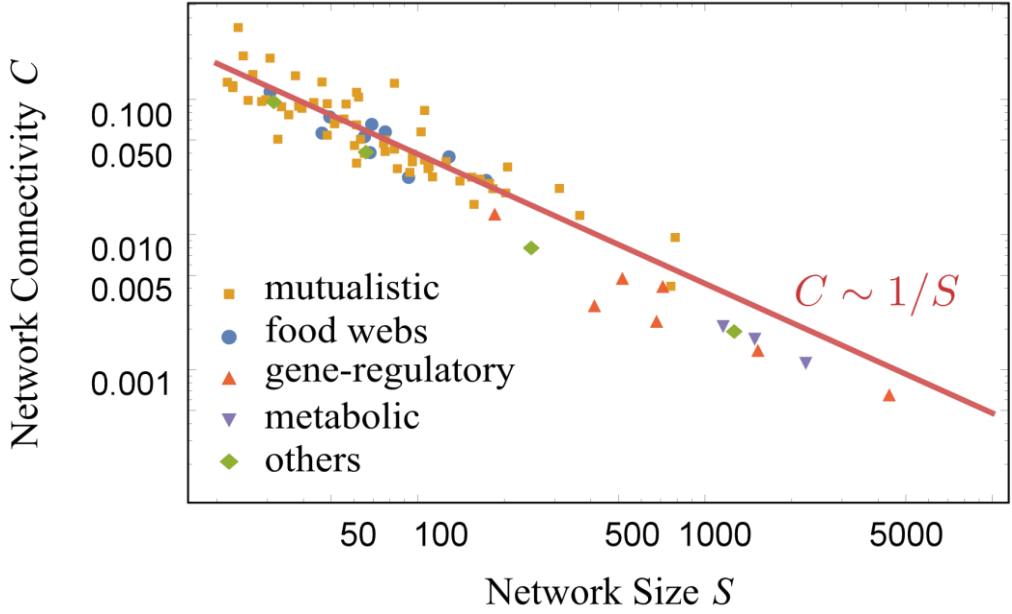

**Figure 1: Sparsity of interactions in living systems**. Connectivity, $C$, is defined as the percentage of active interactions, whereas the system size, $S$, refers to the number of nodes in the graph. We plot $\log C$ as a function of $\log S$ for 83 biological networks (ecological mutualistic communities and food webs[10], gene-regulatory networks[2, 3], metabolic networks[2] and others[2, 3]). Data shows a clear emergent pattern of sparsity in empirical biological networks, as evidenced by the red line $C \sim 1/S$.

**Results**

**Mathematical framework**. We consider a system composed of $S$ nodes (e.g. species, metabolites, genes) characterized by dynamical variables, $\boldsymbol{x} = (x_1, x_2, \ldots x_S)$ (e.g. populations, concentrations, levels of expression) following a generalized Lotka-Volterra dynamics:

$$\dot{x}_i = G_i(x_i) F_i\left(\sum_{j=1}^{S} A_{ij} x_j\right) \qquad i = 1, \ldots S \qquad (1)$$

where $G_i(x) = x$, $F_i(x) = \alpha_i + x$ and the interaction of node $i$ with node $j$ is encoded in the matrix element $A_{ij}$, whose diagonal entries set the scale of the interaction strengths, which we choose equal to $-1$ [21] for the sake of simplicity. We refer to parameters $\alpha_i$ as the growth rates. A non-trivial stationary point of the dynamics Eq.(1), $\boldsymbol{x}^*$, is determined by the interactions within the system, i.e. when $F_i\left(\sum_{j=1}^{S} A_{ij} x_j\right) = 0$, that is $\boldsymbol{x}^* = -\boldsymbol{A}^{-1}\boldsymbol{\alpha}$, and its stability is guaranteed if all the eigenvalues of the Jacobian matrix evaluated at this point, $J_{ij} = x_i A_{ij}$, have a negative real part. Generalized Lotka-Volterra dynamics have used to model the time evolution of ecological systems [21, 22], human microbiota dynamics [23], gene expression [24] and other biological systems [25], where $x_i$ represents the density of the $i$-th species, and therefore we focus on the stable and feasible stationary solutions of the dynamics [21, 18] ($x_i^* > 0$). Notwithstanding this, in the Supplementary Information, we recast all the results for a solution involving more generic non-linear dynamics.

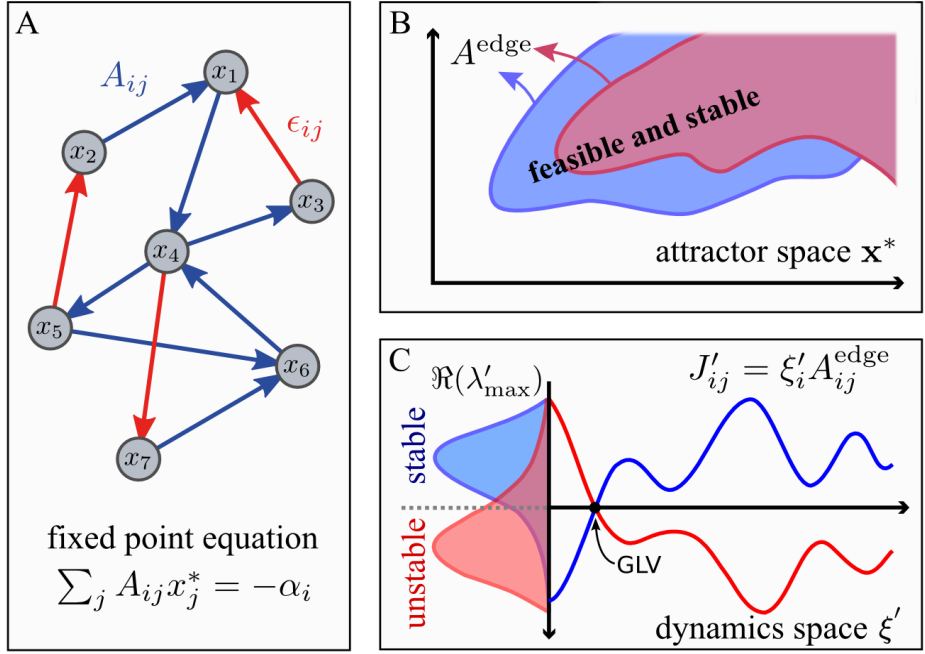

**Figure 2: Measuring explorability and dynamical robustness.** We consider a system of $S$ nodes representing species, genes, metabolites, ... ($S = 7$ in the picture), whose state $\boldsymbol{x} = (x_1, \ldots x_S)$ obeys the GLV dynamics $\dot{x}_i = x_i \left( \alpha_i + \sum_{j=1}^{S} A_{ij} x_j \right)$, with $A_{ij}$ encoding the network structure and strength of interactions. **(Panel A)** We start from a tree-like network with $S$ links, represented by the blue-colored links, and, for such a topology, we search for the feasible and stable fixed points as we vary the interaction strengths. **(Panel B)** The spanned volume sketched by the blue-shaded region (a 2D projection of the $S$-dimensional space) corresponds to the network explorability. **(Panel C)** Starting from an interaction matrix $A_{ij}^{edge}$, whose fixed point of the GLV dynamics is at the edge of stability (black dot), the corresponding dynamics is perturbed and the attractor stability is evaluated in order to test the dynamical robustness of the system. As shown in the Methods, the stability of the modified attractor, as a result of the perturbed dynamics, can be simply encompassed in the Jacobian, $J'_{ij} = \xi'_i A_{ij}$, where $\boldsymbol{\xi}'$ is a random vector. The histogram of the maximum real part of the eigenvalues of $J'_{ij}$ is sketched in Panel C as $\boldsymbol{\xi}'$ is varied. Finally, we increase the connectivity of the network by including additional fixed strengths, $\epsilon_{ij}$ (red edges in the graph), to the network. The same previous analysis is performed again by varying the strengths of the blue links. The corresponding results are shown in red in the panels on the right.

**Explorability**. In order to introduce the concept of explorability, let us consider an interaction network $A_{ij}$ in Eq. (1) for which the strength of the interactions is modifiable but not the who-interacts-with-who (i.e. the topology is fixed). Changing such values one modifies the corresponding attractors as well as their stability. Stated briefly, we define the explorability as the volume in $\mathbb{R}^S$ spanned by all the feasible and stable attractors when one modifies the weights of the matrix elements $A_{ij}$ (see Fig. 2). Notice that, with such a definition, the fully-connected network has the largest possible explorability, since any other topology is attainable by taking some of the matrix entries arbitrarily close to zero. However, might the optimal or quasi-optimal solutions be indeed the ones where most of the interactions are turned off, as suggested by the observational data? Moreover, in the fully connected case, many interaction parameters have to be specified (there are $S(S-1)$ matrix elements that can be varied), and spanning all the possible attractors become a complicated and fine-tuning problem, which does not seem to be a feasible situation in biological systems [26]. Therefore, do sparse networks offer in any way an optimal explorability and consequently a more efficient way to reach new stationary states by just changing few interactions?

To answer these questions, we started by analyzing the extreme case of a sparse topology with just S links, i.e. a tree-like network with connectivity $C = 2/S$ (see Fig. 2) (the factor 2 comes from the fact that we also count the self-interactions). In this case, computing the explorability becomes a much simpler task because of the low number of interactions, and we were able to develop an analytical solution of this problem (see Methods). As a second step, we analyzed the explorability of networks with higher connectivities by adopting the following approach: we introduce additional links to the tree-like topology and then compute the explorability, $V_E(\epsilon)$, by fixing the weights to the value $\epsilon_{ij}$ for any extra link $ij$ (the matrix with these extra links was denoted by $A(\epsilon)$) (see Fig. 2). By sampling different values and locations of the added links, we constructed a histogram of the explorabilities, $P(V_E|C)$, without distinguishing different topologies with the same connectivity (see Methods for technical details). We found, both analytically and numerically (right panel in Fig. 3), that the explorability of the optimal tree-like network is indeed statistically higher than the one for more dense networks and that the average explorability decreases as the connectivity of the interaction network increases (inset of Figure 3).

**Dynamical robustness.** Another crucial property of complex interacting system is their robustness to perturbations [27, 28]. The standard measure of stability (known as asymptotic resilience in ecology [29, 30]), is defined as the capacity of the system, after a perturbation of the stationary state, to return to the original stationary state whereas the dynamics is kept fixed. In this regard, as explained above, we have considered only resilient (and feasible) stationary states. On the other hand we can also study how the stability of the system is modified as a result of a perturbed dynamics, $\dot{x} = (G + \delta G)(F + \delta F)(x)$, where $\delta G$ and $\delta F$ represent the perturbations with respect to the original dynamics. As a consequence of this kind of perturbation, both the original stationary states and their degree of stability are modified. We refer to it as dynamical robustness of the system, in order not to confuse this new measure of stability with the resilience. Indeed, dynamical robustness is the capacity of the system to re-organize after the perturbation so that the new stationary state of the system is close to the original one and still stable. Understanding the role of network architecture in the dynamical robustness of a system with many degrees of freedom is an important challenge since it impacts on our capacity of both preventing system failures and to design more robust networks to tolerate perturbations to the system dynamics. We find the pleasing result that the Jacobian matrix, $J'_{ij}$, at the new stationary state retained the same form as in the original dynamics, that is $J'_{ij} = \xi'_i A_{ij}$, where $\xi'_i$ depends on the specific details of the perturbed dynamic (see Methods and Supplementary Information). We focused on the most relevant instance corresponding to the critical case of interaction matrices with marginally stable attractors (i.e. with $Re(\lambda_{max}) = 0$), that we denoted by $A^{edge}$, and we studied if the stability of such attractors was enhanced ($Re(\lambda'_{max}) < 0$) or diminished ($Re(\lambda'_{max}) > 0$) when the dynamics was perturbed (see panel C in Fig. 2). Since we wanted to keep the analysis as general as possible, in the spirit of a random matrices approach, we randomly generate $\xi'$ from a distribution $P(\xi')$. For each connectivity $C$, we fixed the additional links to $\epsilon_{ij}$ and then looked for the set of matrices $A^{edge}(\epsilon)$. The dynamical robustness of the system as a function of the connectivity of its interaction matrix can be quantified in terms of the distribution, $P(Re(\lambda'_{max})|C)$, of $Re(\lambda'_{max})$ for the perturbed dynamics, by taking different realizations of $\epsilon_{ij}$, and for each one, by generating random values of $\xi'_i$ (see Fig. 2 and Methods for a more specific measure). The right panel of Fig. 3 shows the histogram of $Re(\lambda'_{max})$ for different connectivities. Again it can be seen that the case of tree-like network leads to the best performance, whereby the attractor of the perturbed dynamics generally becomes stable (before the perturbation, the system was marginally stable), that is $Re(\lambda'_{max})$ is negative, and the modulus reaches larger values than in the corresponding case of networks with higher connectivity. Therefore, our analysis shows that sparse tree-like networks have both a larger explorability and larger dynamical robustness than random networks with higher connectivity.

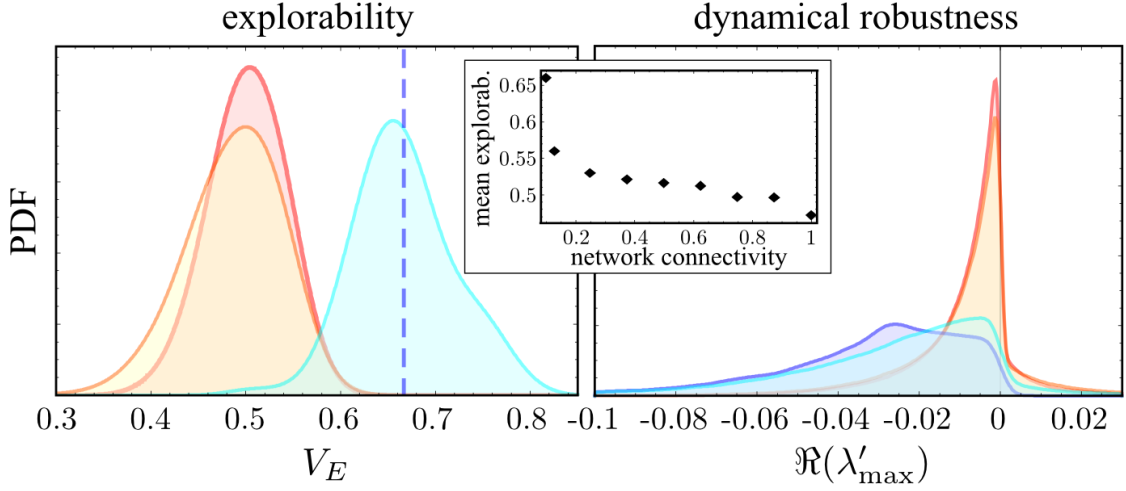

**Figure 3: Explorability and dynamical robustness for different connectivities**. **Left Panel**: Probability distribution functions (PDF) of the explorability $V_E$ for the optimal tree-like graphs and networks with $C = 0.5$ obtained by adding extra links with random (uniformly distributed) locations and weights $\epsilon_{ij}$ taken from a zero-mean Gaussian distribution with standard deviation $\sigma_\epsilon = 0.1$, for a network size $S = 20$. First we compute $V_E$ in the simple setting of uniform concentrations and growth rates, i.e. $x_i^* = x^*$ independent of $i$ and $\alpha_i = 1$ (see Methods). The explorability $V_E = 2/3$ of the tree network (calculated analytically - dashed blue line) is larger than the one corresponding to graphs with higher density (red curves are computed by taking $10^3$ independent realizations of the added links). Similar results hold also in the more general setting of non-uniform concentrations and growth rates ($\sigma_x = \sigma_\alpha = 0.1$, see Methods). Even with such a variability, the tree-like network (cyan curve) generally exhibits higher values of the explorability that more dense networks (orange curve for $C = 0.5$). **Inset**: mean value of the explorability (computed from the PDF of $V_E$) as a function of the connectivity in the homogeneous case ($x_i^* = x^*$ and $\alpha_i = 1$). **Right panel**: Given $A_{ij}^{edge}$ (see Figure 2C), we calculate the Jacobian of the perturbed dynamics $J'_{ij} = \xi'_i A_{ij}^{edge}$. The corresponding eigenvalue with largest real part ($\lambda'_{max}$) gives the degree of stability of the new attractor, from which we measure the dynamical robustness of the system (see Methods). We generate $10^3$ configurations of $\epsilon_{ij}$'s as done for the left panel ($\epsilon_{ij} = 0$ correspond to a tree-like network), and for each one $10^3$ values of $\xi'_i$ (encoding the perturbation of the dynamics) from a uniform distribution in $[0, 1]$, and compute the system response through the distribution $P(\lambda'_{max}|C)$. Both homogenous and non-homogenous settings have been analyzed for the tree-like network and $C = 0.5$ (same color code of the left panel). In all cases, we find that increasing the network connectivity shifts the distribution of $Re(\lambda'_{max})$ towards the less stable region. Qualitatively similar results are obtained for larger values of $\sigma_\epsilon$, $\sigma_x$ and $\sigma_\alpha$ (see Supplementary Information).

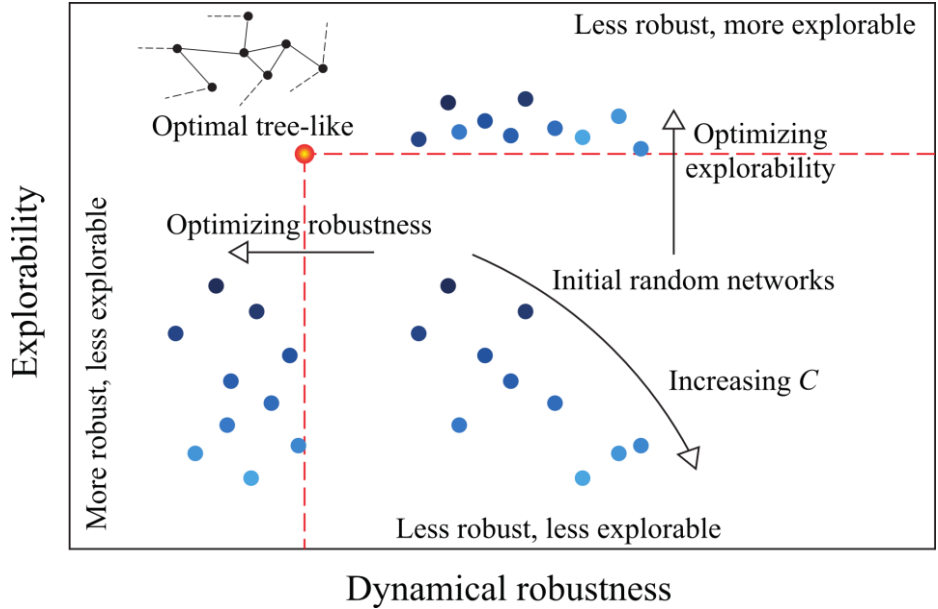

**Figure 4: Sketch of the optimization process** for general networks (details are given in the Supplementary Information). Starting from randomly generated interaction matrices, we fix the topology and optimize either the explorability or the dynamical robustness in the $\epsilon_{ij}$-space. Optimal solutions are found, after extensive search and fine-tuning of the parameters, to be slightly better than the optimal tree-like structure. However, a slightly better explorability than the one of an optimal tree network, generally leads to a worse dynamical robustness, and vice versa. Therefore, sparse topologies statistically offer optimal values of both explorability and dynamical robustness without a fine-tuning of the parameters.

**Optimization approach.** We then went one step further and compared the explorability and dynamical robustness of tree-like networks with graphs constructed via an optimization process rather than randomly generated. To this end, we performed an optimization of the weights of the added links,

$\epsilon_{ij}$, in order to increase the explorability $V_E(\epsilon)$. Similarly, we could also optimize the weights to increase the dynamical robustness (see Methods). The landscape of both the explorability and dynamical robustness appeared to be highly irregular with many local minima when increasing the connectivity (see Supplementary Information). Fig. 4 summarizes the main qualitative results from the optimization process. The explorability reached for networks with connectivities larger than the one for tree-like networks turned out to be very close to the corresponding value for the optimal tree-like topology, but, in general, such networks exhibited low values of the dynamical robustness. Similarly, by optimizing the dynamical robustness, we ended up very close to the value of the optimal tree-like network, but, remarkably, without improving the explorability. In addition, we could simultaneously optimize both the explorability and dynamical robustness, leading to slightly better solutions than the optimal tree-like network, but such a multi-objective optimization process worked only for tree-like networks with one or two additional links, and it was totally inefficient for more dense structures (see Supplementary Information). In conclusion, sparse networks provide optimal values for both the explorability and dynamical robustness without fine-tuning many of the interaction strengths.

**Self-similarity.** Finally we proved that the property of sparsity is self-similar, since on aggregating sparse interacting communities we obtained larger sparse communities (see Supplementary Information for details). For example, joining two networks with a tree like topology using a single link led again to a network with a tree-like topology. Similarly, if sparse networks with $S$ nodes have $aS - b$ links, with $a$ and $b$ integer constants, then joining two such networks with $S$ and $S'$ nodes using $b$ links leads again to a sparse network with $a(S + S') - b$ links. Therefore the optimal features of sparsity are conserved on assembling or disassembling processes, thereby avoiding any drastic change in the stability [31].

**Discussion and conclusions**

Our proposed solution has implication also in the understanding of the relation between stability and complexity in real ecosystems. In fact, our view reconciles previous theoretical arguments, encapsulated in what is called the complexity-stability paradox [32, 33], for which large ecosystems will probably be unstable. The essence of the argument [32, 33] can be summarized as follows. The linearized dynamics for the population density around a stationary state depends on what is known as the community matrix, $M$. If all eigenvalues of $M$ have negative real parts, then the stationary point is also stable against small perturbations of the stationary populations. A null model corresponds to assume that $M$ is a random matrix whose diagonal elements (the self-interactions) are chosen equal to $-d < 0$, whereas the off-diagonal elements are zero with probability $1 - C$ and with probability $C$ are drawn from a probability distribution with zero mean and variance $\sigma^2$. Under this null hypothesis, one finds (see [32, 33] and references therein for rigorous results) that the stationary point is unstable with probability $1$ if $\sigma\sqrt{CS} > d$, where $S$ is the number of species in the ecosystem, a measure of its biodiversity. This result holds if $S$ is assumed to be large enough. Thus if $\sigma$ and $d$ do not have peculiar scaling with the network size, highly complex ecosystems (i.e. with high $CS$) are not stable: a prediction that is in contradiction with empirical data [31, 11, 34]. However if the interaction network is sparse, i.e. $C \sim 1/S$, the above inequality becomes independent of $S$ and the stability of the ecosystem is not threatened by high biodiversities: sparsity in ecological interaction networks allows to have stable large living interacting systems [31, 11, 18]. Recent theoretical findings evidence that an increase of the interconnectivity between multiple systems composed themselves of interacting units can have a strong impact in the vulnerability of the whole system [35]. In the same vein, our results suggest that sparsity is a key feature allowing living systems to be poised in a state that confers both robustness and adaptability (explorability) to best cope with an everchanging environment and to promptly react to a wide range of external stimuli and to resist to perturbations.

Finally, we stress that our results do not depend on the specific details of the system and thus can be applied in many other fields. For example, a possible application might be in the design of artificial learning machines such as deep neural networks [36, 37]. There is a mounting evidence that deep learning often finds solutions with good generalization properties [38, 36] and it has been recently shown [39] that in order to achieve such a good performance it is crucial to have regions of the optimization landscape that are both robust and accessible, independently on the particular task or of the training dataset. On the other hand, maximization of computation efficiency is a crucial point when designing them: dee networks are very dense, as each node is connected to all other nodes of the adjacent layers [37], which makes multi-layer neural networks computationally hard to train. Our solution suggests that designing sparse neural networks will increase exploitability of the system while, at the same time, would improve the convergence and robustness properties of the existing optimization algorithms.

## Methods

*Exploring the space of attractors*: The framework introduced in the main text requires to explore the $S$-dimensional space of attractors. For this reason, we introduce various degree of approximations in our setting by first reducing the analysis to the subspace of attractors with homogeneous components, i.e. to the bisector $x_i^* = x$. Equivalently, we considered the simplest uni-parametric case $\alpha_i = \alpha$ (that we can fix without loss of generality to $\alpha = 1$). Even thus, the approximation looks, *a priori*, rough but it leads to rather reasonable estimates of the sparsity and of the dynamical robustness associated with a given network topology. Indeed, by introducing some heterogeneity in the components of $\boldsymbol{x}^*$ and $\boldsymbol{\alpha}$, we found that the results remained, at least qualitatively, the same as the homogeneous case.

*Measuring explorability*: The explorability of a tree-like network with $S$ links ($C = 2/S$) can be found by studying the following inverse problem: by fixing the parameters $\boldsymbol{\alpha}$ and moving along the space of attractors, one can retrieve the non-zero $S$ values of $A_{ij}$ according to the fixed point equation $\sum_{j=1}^{S} A_{ij} x_j = -\alpha_i$, and this can then be used to check the stability of the associated attractor $\boldsymbol{x}^*$. The solution exists if for each node, $i$, there is at least a node $j$, such that $A_{ij} \neq 0$ (see Supplementary Information). The same procedure can be applied to the more general case where extra links are added to the network, each one with a given strengths $\epsilon_{ij}$ (see Supplementary Information for more detail).

In the simplest situation in which $x_i^* = x^*$ and $\alpha_i = \alpha$, we observed that, in almost all the cases (100% for tree-like networks and, e.g., more than 98% for $C = 0.5$ and $\sigma_\epsilon = 0.1$), $Re(\lambda_{max})$ was positive for a small $x^*$ and negative for a large $x^*$, intersecting $Re(\lambda_{max}) = 0$ at a single value $x_c^*$ (see Supplementary Information for detail). Therefore, $V_E = L - x_c^*$, where L is a sufficiently large constant, is a good definition of the explorability. Although such volumes could be infinite, we are always interested in comparing them for different topologies. In this simple setting, we can compute analytically the explorability of a tree-like network with $S$ links (see Supplementary Information), finding that, among all the possible tree-like topologies, the one with just a loop composed by 3 nodes leads to the optimal explorability. This structure (which we refer to as the optimal tree-like network) constitutes our reference network when increasing the connectivity.

*Increasing heterogeneity*: We enlarged the explored region of attractors by sampling the space around the bisector $x_i = x^* + p_i$, where $p_i$ is a Gaussian random variables with a zero mean and standard deviation $\sigma_x$. Similarly, we can introduce variability in the model parameters $\alpha_i = \alpha + q_i$ (same distribution as $p_i$ with the standard deviation $\sigma_\alpha$). In this case, we counted *all* the attractors at the edge of stability (within a small error $|\lambda_{max}| < 10^{-2}$), and for each one we evaluated $V_E = 1 - \sum_{j=1}^{S} x_i^* / S$ as the most straightforward generalization of our previous definition of $V_E$. Curves in Fig. 3 for the case with heterogeneity were obtained using $10^2$ independent realizations of $\epsilon_{ij}$, and, for each one, 10 realizations of $(p_i, q_i)$.

*Measure of dynamical robustness*: In order to define a measure of stability of an interaction matrix with respect to perturbations of the dynamics, we proceeded as follows. For a given choice of the added links $\epsilon_{ij}$'s, we sought for the distribution $P(Re(\lambda'_{max})|C)$ of the perturbed dynamics, and we took the value located at the 5th percentile as an indicator of how much stability could be gained under the perturbations of the dynamics (see Supplementary Information for other choices of this measure).

**Acknowledgements**   We thanks Jayanth Banavar for inspiring discussions and suggestions.

**Author Contributions**   S.S., J.H. and A.M. designed the research, D.M.B. carried out analytical and numerical calculations. All the authors did a critical revision of the work and contributed to other aspects of the paper.